\pdfoutput=1

\documentclass[sigconf]{acmart}

\makeatletter                   
\def\mdseries@tt{m}             
\makeatother                    
\usepackage[plain]{fancyref}
\usepackage[draft=true]{minted} 
\usepackage{color}
\usepackage{hyperref}           
\hypersetup{
    colorlinks=true,
    linkcolor=blue,
    filecolor=red,      
    urlcolor=magenta,
    breaklinks=true,            
}
\usepackage{breakurl}           
\setcopyright{none}

\def\acmBooktitle#1{\gdef\@acmBooktitle{#1}}
\acmBooktitle{Proceedings of \acmConference@name
       \ifx\acmConference@name\acmConference@shortname\else
         \ (\acmConference@shortname)\fi}
		 
\usepackage{booktabs}
\usepackage{makecell}

\usepackage{balance}
\usepackage{array}
\usepackage{multirow}
\newcommand{\PreserveBackslash}[1]{\let\temp=\\#1\let\\=\temp}
\newcolumntype{C}[1]{>{\PreserveBackslash\centering}m{#1}}
\newcolumntype{R}[1]{>{\PreserveBackslash\raggedleft}m{#1}}
\newcolumntype{L}[1]{>{\PreserveBackslash\raggedright}m{#1}}

\AtBeginDocument{%
  \providecommand\BibTeX{{%
    \normalfont B\kern-0.5em{\scshape i\kern-0.25em b}\kern-0.8em\TeX}}}

\begin{document}

\copyrightyear{2020}
\acmYear{2020}
\acmConference[WWW '20 Companion]{Companion Proceedings of the Web Conference 2020}{April 20--24, 2020}{Taipei, Taiwan}
\acmBooktitle{Companion Proceedings of the Web Conference 2020 (WWW '20 Companion), April 20--24, 2020, Taipei, Taiwan}
\acmPrice{}
\acmDOI{10.1145/3366424.3383570}
\acmISBN{978-1-4503-7024-0/20/04}

\settopmatter{printacmref=true}

\title[Layered Graph Embedding for Entity Recommendation]{Layered Graph Embedding for Entity Recommendation using Wikipedia in the Yahoo! Knowledge Graph}

\author{Chien-Chun Ni}
\email{cni02@verizonmedia.com}
\affiliation{%
  \institution{Yahoo! Research}
}

\author{Kin Sum Liu}
\email{kiliu@cs.stonybrook.edu}
\affiliation{%
  \institution{Stony Brook University}
}

\author{Nicolas Torzec}
\email{torzecn@verizonmedia.com}
\affiliation{%
  \institution{Yahoo! Research}
}

\begin{abstract}
In this paper, we describe an embedding-based entity recommendation framework for Wikipedia that organizes Wikipedia into a collection of graphs layered on top of each others, learns complementary entity representations from their topology and content, and combines them with a lightweight learning-to-rank approach to recommend related entities on Wikipedia. Through offline and online evaluations, we show that the resulting embeddings and recommendations perform well in terms of quality and user engagement. Balancing simplicity and quality, this framework provides default entity recommendations for English and other languages in the Yahoo! Knowledge Graph, which Wikipedia is a core subset of.


\end{abstract}


\maketitle

\section{Introduction}

\begin{figure}[hbp]
    \centering
    \includegraphics[width=0.9\columnwidth]{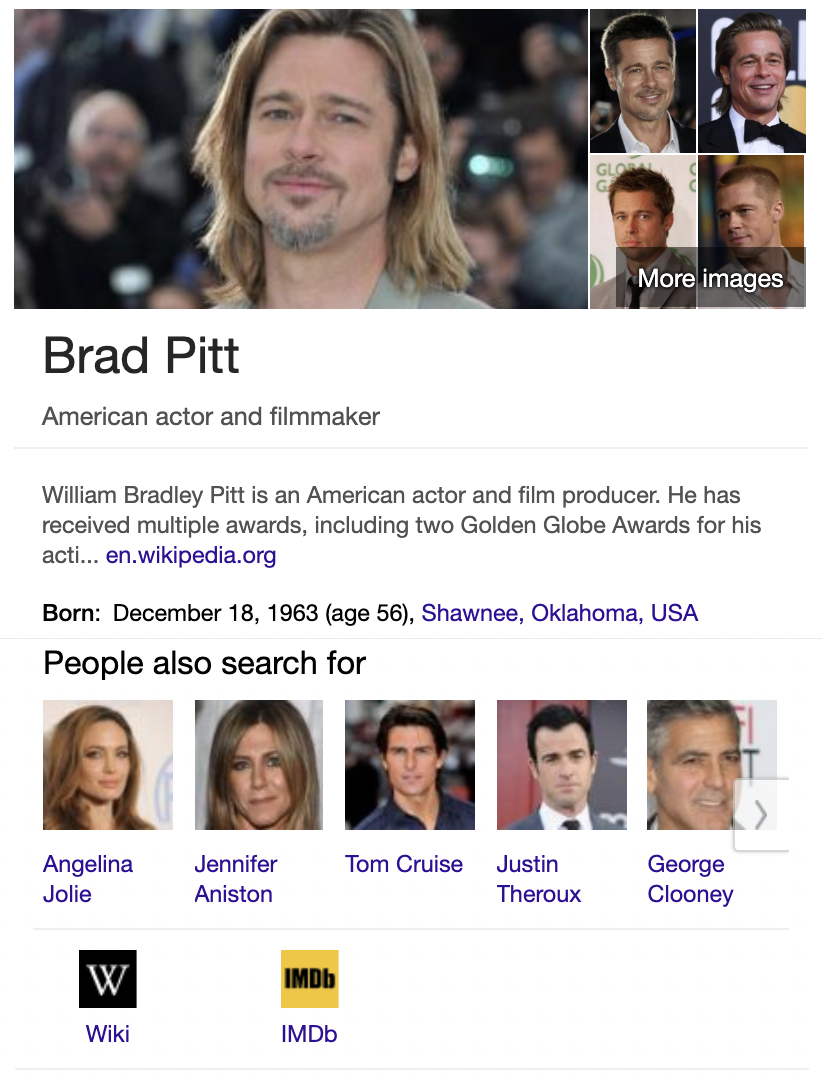}
    \caption{Knowledge panel with entity recommendation for the query ``Brad Pitt'' on Yahoo! Search in the U.S.}
    \label{fig:kg_sample}
\end{figure}

Entity recommendation is the problem of suggesting a contextually-relevant list of entities in a particular context. Here, we focus on the problem of recommending relevant related entities given a particular entity as input, independently of the user. This is a fundamental problem for Web search engines such as Google, Bing or Yahoo! Search, which typically displays a knowledge panel with factual information and a list of related entities relevant to the searched entities on their search result pages. On web-scale, the problem is even more challenging given the size of the massive search space, various subjectivity of the task, lack of clearly defined context, and transient nature of some of the relationships, etc. See Figure~\ref{fig:kg_sample} for an example of knowledge panel with related entities recommendation for the query ``Brad Pitt'' on Yahoo! Search. 

In this context, we present a framework for recommending related Wikipedia entities given a Wikipedia entity as input. Considering Wikipedia as an organized collection of information about entities, this framework first organizes and combines Wikipedia into a unified collection of \emph{layered graphs} placed on top of each others, with Wikipedia pages in various languages as vertices, their textual content as documents attached to each vertex, and the various types of links between them as edges. Then, our framework learns complementary entity representations based on the topology and content of the unified graph, and uses the resulting embeddings to generate candidates for entity recommendation with the assumption that entities related or similar in Wikipedia would be relevant entities to recommend. Finally, it ranks those candidates using a lightweight learning-to-rank approach to fine-tune the recommendations.

The overall approach is user-independent, works across many languages, and is agnostic to entity types. Beyond ``companies most related to a given company'', one can also ask for the ``persons most related to a given company'' or the ``books most related to a given place''. Through offline evaluations, we show that this approach generates results that are significantly better than using document embeddings alone in terms of embedding quality and candidate generation coverage, and better than Google in terms of entity recommendation accuracy for tasks such as recommending the most related companies for a given company\footnote{The datasets are available at \url{https://webscope.sandbox.yahoo.com/}}. Through online evaluations on live Web Search traffic, we also show that this approach performs well in terms of user engagement. Balancing simplicity and quality, it now generates default entity recommendations for English and other languages in the Yahoo! Knowledge Graph, which Wikipedia is a core subset of. 

The rest of the paper is organized as follows. Section 2 introduces related works to provide context. Section 3 formalizes the problem and describes the different components of our recommendation framework, focusing on how we organize Wikipedia into a collection of layered graphs, how we learn topology-based and document-based entity representations from them to generate candidates for entity recommendation, and how we train and use a lightweight ranking model on top of them to fine-tune the final recommendations. Section 4 evaluates the quality and performance of our approach, summarizing the datasets, methodologies, and experimental results that we used and generated during our offline and online experiments. Section 5 concludes the paper.

\section{Related Works}




\subsection{Entity Recommendation}
Entity recommendation enhances the utility of the search engine when answering information seeking queries. 
When users query entities, knowledge panels next to the search result help users to discover related knowledge\cite{henry2016providing}.
However, current knowledge bases may contain millions and even billions of entities. Reasoning the relatedness among this huge pool of entities is difficult. Multiple efforts \cite{blanco2013entity, kang2015learning, yu2014building, bi2015learning} have been carried out to rank the related entities for recommendations. By considering the knowledge base as an entity-relationship graph, they rank only explicitly related entities and design features on this graph. Although proved to be effective in practice, they require heuristics for candidate discovery and heavy feature engineering for entity ranking in each different domain. On the other hand, Wikipedia contains a rich source of textual information about entities as a collection of articles with hyperlinks between them. In \cite{aggarwal2015leveraging}, authors designed features using Wikipedia to rank related entities. This approach only considers (co)-occurrence so does not take advantage of the global topology of the Wikipedia graph. In this work, we propose a general framework to leverage the geometry of Wikipedia by learning the embedding on different layers and decompose entity recommendations into candidate suggestion and entity ranking. This allows ranking diversified candidates in an efficient manner.


\subsection{Graph Embedding}

Following word2vec's \cite{mikolov2013efficient} success of learning word representation from a large corpus of unstructured text, word2vec has been extended to multiple topics such as document embeddings (doc2vec) \cite{le2014distributed} and graph embeddings \cite{Cai2018-nl, Goyal2018-gf}. DeepWalk \cite{perozzi2014deepwalk} generalizes the idea to graph data by embedding the graph's vertices in a continuous latent space. This representation is also trained in an unsupervised way with the hypothesis that vertices visited by the same random walks within a short window should be closer to each other in the embedding space. Specifically, the authors focus on social networks and leverage the social representation encoded by the embedding for other supervised classification tasks. node2vec \cite{grover2016node2vec} extends DeepWalk by relaxing the definition of neighborhoods of vertices. It proposes a biased random walk procedure that introduces a search bias that controls the exploration of the walks within the graph. This flexibility allows the effective learning of different kinds of similarities between vertices.

Recently graph embedding techniques, especially the random walk based ones, have been extended to the multi-layered graph. In \cite{Liu2017-ly}, for a walk within a layer, a parameter is given to determine if the walk will ``jump'' to different layers.  In \cite{Song2018-mo}, the embedding of a multi-layered graph is learned from creating extra inter-layer edges inferred by the similarity of local neighborhood structures. \cite{Bagavathi2019-ry} proposed various methods of doing random walks across different layers and directly take the union of these walks together for embedding training. In \cite{Zhang2018-to}, a multiplex network embedding consists of a common embedding shared across layers and layer-specific embeddings. Besides random walk based methods, in \cite{Li2018-nf}, a spectral based multi-layered heterogeneous embedding is proposed, in this method, within-layer edges and cross-layer edges are trained separately and merged. Creating inter-layer edges to perform random walks across layers requires heuristics for edge construction and imposes computing constraints when dealing with large graphs. Our approach simply performs independent walks on different layers and infer cross-layer relationships by mapping the entities in the same embedding space. It also allows incremental addition of new layers which is useful for fine-tuning the embedding for different applications in the real-world setting.

\subsection{The Yahoo! Knowledge Graph}

Launched in 2013 to support entity-oriented services and applications at Yahoo! (now part of Verizon Media), the Yahoo! Knowledge Graph is an organized collection of normalized information about entities modeled as a graph \cite{torzec2014semtech}. Information is extracted from the Web and various public and commercial databases (incl. Wikipedia and Wikidata), and integrated with a central knowledge base where it is normalized (i.e. schema alignment and data normalization), reconciled (i.e. matched and merged \cite{Bellare2013-mv}), and refined (e.g. entity type classification) automatically. Overall, the Yahoo! Knowledge Graph consists of almost a billion entities and billions of facts. It mostly focuses on notable entities and the factual information associated with them: movies, TV shows, music, sports, people, organizations, places, events, politics, health, products, etc. It is accessible via Search and Graph APIs across Verizon Media, to display information about entities, understand queries and longer documents, answer natural language questions, relate and recommend entities \cite{kang2015learning, blanco2013entity}, and other entity-based services.
\section{Methods}
\begin{figure}[]
    \centering
    \includegraphics[width=1\columnwidth]{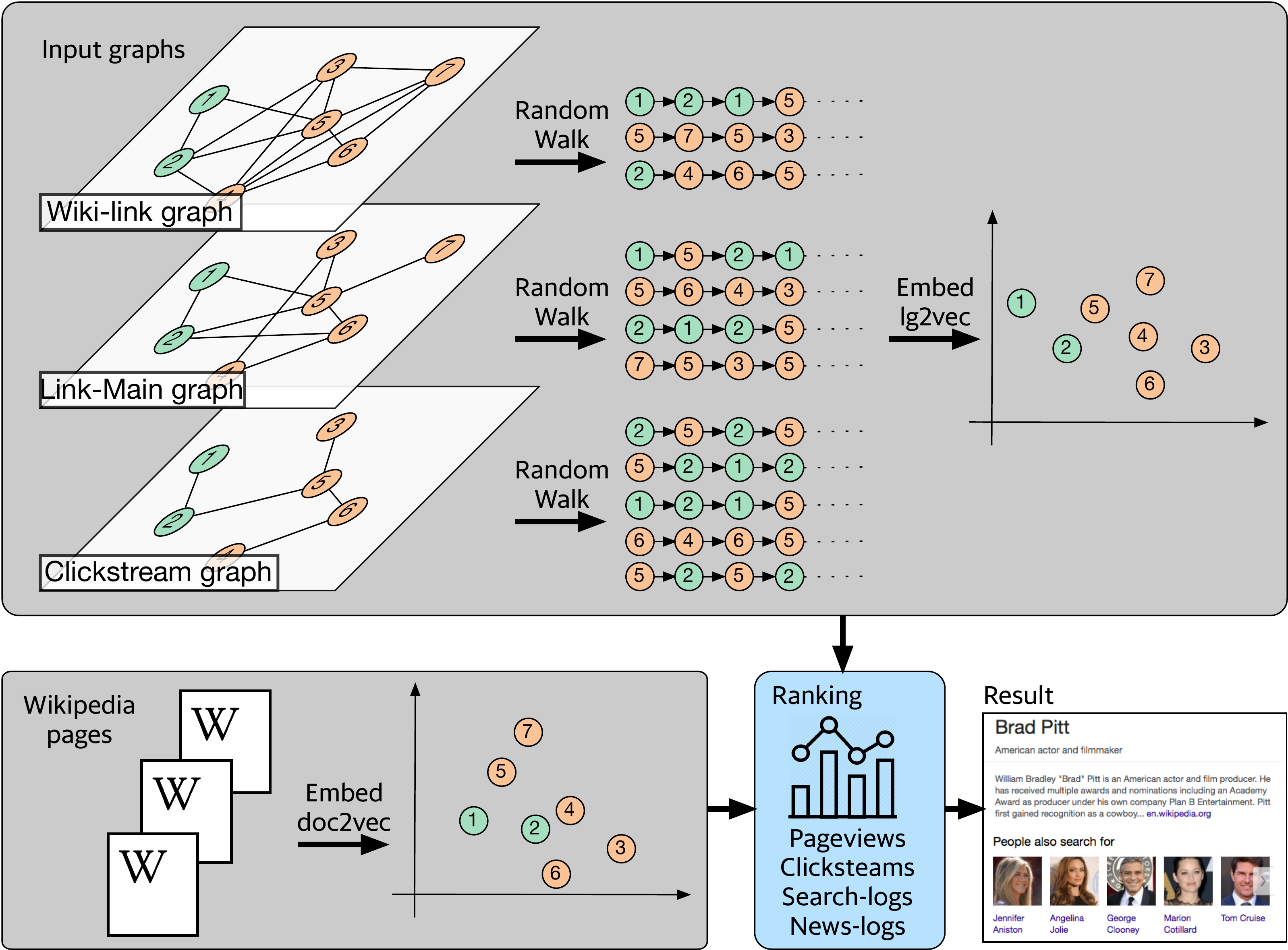}
    \caption{The two-stage workflow: first we construct lg2vec and doc2vec embeddings for candidate generation, then apply the ranker with embeddings similarities and other features for entity ranking, and acquire the final entity recommendations.}
    \label{fig:workflow}
\end{figure}

In this section, we discuss the framework of our method. We first introduce the problem definition and the topology based layered graph embedding lg2vec, then we combined this embedding with the semantic-based embedding doc2vec for candidate generation. Finally, these embedding similarities are aggregated with other features such as node and edge popularities to train the ranking model by pairwise learning-to-rank approach. The workflow is illustrated in Figure~\ref{fig:workflow}.

\subsection{Problem Definition}
Let $D$ be the universe of entities. For a query $q$ which can be an entity or a string that can be mapped to an entity through entity recognition, a partial ordering $D_1 \succ D_2 \succ \ldots \succ D_l$ where $D_i \subset D$ induces a partial ranking: for entities in the same subset, they are ranked the same for the query $q$ and an entity in $D_i$ is ranked higher than an entity in $D_j$ if $i < j$. This ordering can be equivalently expressed by associating a label $y$ from a label set $Y = \{1,2,,\ldots,l\}$ for each entity $d \in D$ such that $l \succ l-1 \succ \ldots \succ 1$. The list of labels for the universe $D$ forms the vector $\boldsymbol{y} \in Y^{|D|}$. Therefore, the training data consists of multiple queries and their associated orderings as $\{(q_1, \boldsymbol{y_1}), (q_2, \boldsymbol{y_2}), \ldots \}$

In the entity recommendation, we aim to train a per-entity scoring model $h(q, d)$ that assigns a score for a pair of query entity $q$ and candidate entity $d$. If we apply this scoring function to every candidate in the universe $D$, then we again get a list of scores $\boldsymbol{h}$ so that we can evaluate the performance of this scoring function by computing standard metrics using the ground truth labels $\boldsymbol{y}$.

In our case, our training golden set only partitions the universe into two subsets for each query $q$: relevant $D_r$ and irrelevant $D_{\bar{r}}$ with $D_r \succ D_{\bar{r}}$ so the label is binary $(|Y| = 2)$. And we decompose the entity recommendation into two stages: candidate generation and entity ranking. In candidate generation, we want to narrow down the universe into only a set of candidates $N(q)$ for the query $q$ with $|N(q)| << |D|$. The performance of this stage can then be measured by the coverage of this candidate set: $|N(q) \cap D_r|/|D_r|$. In entity ranking, we learn a scoring function $h$ and apply it to entities in $N(q)$ to rank and then sort to get an ordered list. Then we can measure the performance of this ranking by Normalized Discounted Cumulative Gain or Mean Average Precision.

\subsection{The Layered Graph}

Knowledge graphs, such as the graph formed by facts or links on Wikipedia, are usually directed multi-graphs where each vertex represents an entity (a Wikipedia page) and each (directional) edge represents a relationship. In a knowledge graph $G=(V,E,W)$, most of the input data are represented in the form of entity relationship pairs $(v_1,v_2,r,w)$ where $v_1, v_2 \in V$ are related entities with relationship $r \in E$ with weight $w \in W$ if an edge weight is given.

A layered graph $G_L = (G_1, G_2, \ldots, G_m)$ is a master knowledge graph that is composed by different subgraphs, where $V_L=\cup_{i=1}^m V_i, E_L=\cup_{i=1}^m E_i$. These subgraphs might be constructed under different purposes, complementary to the others, but are all in one entity space. For example, while taking all pages and possible links in Wikipedia as the master graph, each individual language’s Wikipedia can be treated as one complimentary subgraph of the master Wikipedia graph. Similarly, a popularity based information such as clicks for page links, can also be easily transformed into entity relationship pairs as the other subgraphs of the master Wikipedia graph. This layered graph structure enables us to easily plug in new datasets. For example, the base embedding with general knowledge from Wikipedia can be enhanced to a finance-oriented embedding by inserting the stock and trading related data layer.





\subsection{lg2vec with Layered Graph}

The construction of the lg2vec embedding for the layered graph can be seen as an extended version of DeepWalk. For a collection of graph layers $G_L = (G_1,G_2,\ldots,G_m)$, we construct the corpus $C = (C_1,C_2,\ldots,C_m)$ via biased random walks on the graph layers. For each layer $G_k$, $C_k$ is constructed by initiating $n_k$ biased random walks with $10$ hops length from each node in the graph. Here, the outgoing probability of a biased random walk is proportional to the outgoing edge weight. 

The benefit of constructing corpora on the layered graph structure instead of on the union graph is the freedom to adjust the ``preference'' of each layer for the embedding training process. By choosing the number of random walks on each layer $n_k$ as a knob for corpus construction, we can better parameterize the graph embedding in a more fine-grained way rather than binarily pick a subgraph to union. For example, if we want to tailor an embedding that is specific for recommendations in the French market, we can simply prioritize the FR-Wiki layer by performing more random walks on it. The layered graph structure can also play the role of a bootstrap. Suppose we receive new data from Yahoo! Sport or Yahoo! Finance that are graph datasets on their specific domains, instead of directly learning new embeddings from these graphs and losing all other general relationship learned from Wikipedia, we can simply regard these new datasets as new layers, construct new corpora from them, and update the existing embeddings. 

After constructing the corpus $C$ from random walks on graph layers, we follow the generic word2vec process to apply a window size (possibly different values for different layers) to define neighborhoods and build the training data following the Skip-gram model. The intuition of Skip-gram in word2vec is that the current node is used to predict the surrounding nodes within the same neighborhood. As a result, two nodes close in graph topology will co-occur in the corpus, and hence close in embedding space.


\subsection{word2vec/doc2vec}
Let the (input and output) representations of the vertices to be $f_I, f_O: V \mapsto \mathbb{R}^d$, where $d$ is the dimension of the embedding space. The vertices' representations can be learned from corpus $C$ by maximizing the log probability of observing the corpus. Following the Skip-gram formulation \cite{mikolov2013efficient}, the optimization objective is defined as

$$ \max_{f_I, f_O} \sum_{(v_I,v_O) \in \text{Skipgrams}(C)} \log \Pr(v_O | v_I)$$
where the probability is modeled as

$$\Pr(v_O | v_I) = \frac{\exp(f_O(v_O)\cdot f_I(v_I))}{\sum_{v\in V} \exp(f_O(v)\cdot f_I(v_I))}.$$
 Since the normalizing factor is difficult to compute, Negative Sampling \cite{mikolov2013efficient, stergiou2017distributed} is used to approximate this log probability by

$$ \log \sigma(f_O(v_O)\cdot f_I(v_I)) + \sum_{v \in V_{neg}} \log \sigma(-f_O(v)\cdot f_I(v_I))$$
where $\sigma(x) = \frac{1}{1+e^{-x}}$. So the learning procedure can be summarized as the iterative updates of the representation using the gradient of the objective that is computed by sampling negative examples for each skip-gram.
doc2vec is a variant of the word2vec model that learns the semantic representations of document tags along with the words. In order to do so, while constructing the corpus, the document tags are added to each corpus, then the word2vec process is applied for these corpora to learn the representations of document tags. In this paper, we use the in-house implementation of the distributed word2vec training system mentioned in  \cite{stergiou2017distributed} on Hadoop as our word2vec training engine.

The lg2vec and doc2vec embeddings provide us an intuitive way to generate candidates for related entities. Since similar entities are embedded in closer positions in the constructed embedding space, we can simply take the k-nearest neighbors of an entity in lg2vec and doc2vec embeddings as candidates. In this way, we can cover candidates with both topological similarities and semantic similarities.

\subsection{Learning-to-Rank}
Besides candidate suggestions, the embeddings also provide pairwise entity similarities as Euclidean distance or cosine similarity. These similarities are perfect features for learning-to-rank (LTR) algorithms. Given a pair of candidate entities, the pairwise loss of LTR tries to learn a binary classifier (ranker) that can decide the better entity with a score given to each entity. The ultimate goal of this ranker is to minimize the inversions in the ranking process, that is, the ranked results are in the wrong order compared with the ground truth. Practically, the score given by the ranker is directly used to rank the result. In this paper, our ranker is learned by the industry-standard gradient boosted tree framework XGBoost \cite{chen2016xgboost}.

\section{Evaluations}

In this section, we evaluate the embedding quality and the performance of our method for candidate generation and entity ranking. We use Wikipedia (a core subset of the Yahoo! Knowledge Graph) as input for the framework, and two entity recommendation datasets as ground truth: one created by Yahoo!'s in-house editorial team and another one crawled from Google's search result pages. The highlights of our evaluations are as follows:

\begin{enumerate}
    \item Regarding embedding quality, the lg2vec embedding yields up to $6\%$ better performance than the doc2vec embedding, and the embedding qualities are shown to be robust with training parameters.
    
    \item Regarding candidate generation, the lg2vec embedding covers up to $88\%$ of the items in the recommendation golden set while picking top $500$ nearest neighbors as suggestions. In comparison, the doc2vec embedding only covers $67\%$.
    
    \item Regarding entity ranking, our lightweight learning-to-rank approach over lg2vec embedding performs $4\%$ better than Google on the ``most related \emph{company} given a \emph{company}'' task when measured against our editorial golden set. Besides, our method can also generate type-agnostic entity recommendations such as ``most related \emph{products} given a \emph{company}'' or ``most related \emph{persons} given an \emph{event}''.
    
    \item Regarding user engagement on Yahoo! Search, our approach increases the click-through rate (CTR) by $1-3\%$ and the number of clicks per session by $500\%$ on the ``most related \emph{company} given a \emph{company}'' task. 
    
\end{enumerate}

\subsection{Datasets}

\subsubsection{Input Graphs}

In order to generate the layered graph, we use the data dumps from Wikimedia\footnote{\url{https://dumps.wikimedia.org}} to build different graphs (layers) based on languages and link types. The statistics for each layer are summarized in Table \ref{tbl:graph_dataset}.

\begin{table}[h!]
    \caption{Statistics of Wikipedia graph datasets}
    \begin{tabular}{ L{3.2cm}  c c }
     \toprule
     Graph (20191201) & \#Nodes & \#Edges \\ 
     \midrule
     En-UniqLink-All & 5,976,815 & 483,819,490 \\
     En-Link-Main       & 5,961,724 & 115,955,119 \\
     En-Clickstream     & 3,818,249 & 35,271,863 \\
     Top10-Combined      & 15,400,279 & 1,227,841,141 \\
     En-Master \footnotesize{(yearly pageview $>$ 8000)} & 1,315,909 & 138,540,067 \\
     Top10-Combined \footnotesize{(yearly pageview $>$ 8000)} & 1,593,252 & 266,461,223 \\
     \bottomrule
    \end{tabular}
    \label{tbl:graph_dataset}
\end{table}

\paragraph{Language-specific layered graphs} The Wikipedia EN link graph (En-UniqLink-All) is built from the \emph{enwiki-latest-pagelinks.sql.gz} dump: each node represents a page in the English Wikipedia and each edge represents a hyperlink from one page to another page in the same Wikipedia. In this graph, all types of links are used (e.g. reference links, category links, template links, etc.) but multi-edges are replaced with simple edges. Also, pages with redirects are transitively processed such that only the target pages are used as nodes and in edges. To enhance the descriptive power of the Wikipedia EN link graph, we build two additional weighted graphs. The Wikipedia EN link-maintext graph is also built from the \emph{enwiki-latest-pagelinks.sql.gz} dump but we only consider links from/to the main body of each article as edges, and the numbers of links between two nodes are treated as edge weight. Other types of links (i.e. category links, reference links, template links, etc.) are considered less relevant as related entities. The Wikipedia EN clickstream graph (En-Clickstream) is built by aggregating monthly counts from the \emph{clickstream-enwiki-*.tsv.gz} dumps. In this graph, the edge weights represent the number of clicks from one page to another over a period of 12 months. Finally, a master graph (En-Wiki-Master) with three layers (i.e. link, link-maintext, and clickstreams) is created by combining the various graphs built from the English Wikipedia.

\paragraph{Multi-lingual layered graph} The process for English is replicated for each of the ten largest Wikipedias (i.e. DE, EN, ES, FR, IT, JA, PL, PT, RU, and ZH), and a master graph is created for each language. From there, a combined link graph is created by combining all the master graphs together. Nodes are union-ed and consolidated across languages using the inter-language links and identifiers in Wikidata\footnote{\url{https://www.wikidata.org}}. Edges are also union-ed, and weighted by summing up the number of occurrences in each language. Edges that only appear in one language are filtered out, assuming they would be less important. The resulting combined graph not only doubles the number of edges in the original graphs but also enable it to favor links and edges from specific languages for different markets. For clarity, in the following section entities are presented with their English Wikipedia IDs.

\paragraph{De-noising} Since Wikipedia contains many empty and noisy pages, we filter out pages whose yearly page view counts is less than 8,000 occurrences combined in the top ten languages. The final combined graph contains approximately 1.6m nodes and 266m edges, and also has three layers: link, link-maintext, and clickstreams.

\subsubsection{Input Documents}

In order to generate the document embeddings, we extract the text of the Wikipedia pages from the latest \emph{*wiki-*-pages-articles-*.xml.bz2} dumps using wikiextractor\footnote{\url{https://github.com/attardi/wikiextractor}}. Since natural language preprocessing varies by language, we only use document embeddings generated from the English Wikipedia in this paper. 

\subsubsection{Golden Sets for Evaluation}

\paragraph{Embedding Quality}
To evaluate the quality of the embedding, we applied the Wikipedia triplets golden set used by doc2vec for quality check in Dai \textit{et al.} \cite{dai2014document}. In this evaluation data, entity triplets $(a,b,c)$ are provided in a form that $(a,b)$ should be more similar than $(a,c)$. This dataset includes $172$ hand-built and $19996$ automatically generated triplets. 
The automatically generated triplets are created by using the category type in Wikipedia with the assumption that one entity should be more similar to the entity in the same category rather than other entities that are several category levels away. Notice that since we filter the Wiki pages by its yearly pageviews, we end up with $170$ hand-built and $2917$ generated triplets fully recognized by our embedding.

\paragraph{Candidate Generation \& Entity Ranking}
We prepared a crawled and an editorial golden set to analyze the performance of candidate generation and entity ranking tasks. For the crawled golden set, we crawled Google knowledge card's ``People also search for'' section for entities that are typed as \emph{Person} or \emph{Company} by YK. The crawling lists for persons and companies are biased sampled by entity's yearly pageview (which implies popularity) from all \emph{Person} and \emph{Company} entities in Wikipedia. The crawl was completed in Dec 2018 with more than $20,000$ results for \emph{Person} and \emph{Company} recommendation golden set. See Table~\ref{tbl:goldensets_stat} for more details.

\begin{table}[ht]
\caption{Statistics of golden sets for entity recommendation}
\resizebox{1\columnwidth}{!}{
\begin{tabular}{lccccc}
\toprule
                      & \multirow{2}{*}{\#Entities} & \multicolumn{4}{l}{\#Recommended Entities} \\ \cmidrule(r){3-6}
                      &                           & Mean        & Min     & Median     & Max    \\ 
    \midrule
    Person (Google)     & 24304                     & 21.55     & 3       & 23    & 97     \\
    Company (Google)    & 21216                     & 14.30     & 2       & 14    & 24     \\
    Company (Editorial) & 1012                      & 9.96      & 6       & 10    & 18     \\
\bottomrule
\end{tabular}
}
\label{tbl:goldensets_stat}
\end{table}

For the editorial golden set, our in-house editors provided recommendation results for \emph{Company} entities that are sampled with weights on yearly pageviews. For each entity, the editor chose at least three ``highly related'' and at least three ``somewhat related'' entities using domain knowledge. The absolute recommendation order in this golden set is not crucial since it is very hard to decide the absolute order of the ``highly related'' entities such as suggesting Samsung or Google first for Apple Inc.. To align this editorial golden set to the same format of the crawled one, we attach the ``somewhat related'' list to the end of ``highly related'' list as one ordered recommendation list. As a result, we have a golden set with $1012$ \emph{Company} suggestions with at least $6$ recommendations. Examples of this editorial golden set are shown in Table~\ref{tbl:editoral_goldenset}.

\begin{table}[ht]
\small
\caption{Examples for editorial golden set on \emph{Company}}
\resizebox{1\columnwidth}{!}{
    \begin{tabular}{L{0.5cm}C{3.8cm}C{3.9cm}}
    \toprule
    Entity                      & Highly related                & Somewhat related                  \\
    \midrule
    \multirow{4}{*}{Pixar}      & DreamWorks Animation          & Illumination (animation company)  \\ 
                                & Sony Pictures Animation       & Blue Sky Studios                  \\ 
                                & Walt Disney Pictures          & Warner Bros.                      \\ 
                                & Walt Disney Animation Studios & Marvel Comics                     \\ 
    \midrule
    \multirow{5}{*}{Apple Inc.} & Samsung                       & Amazon (company)                  \\
                                & Google                        & HP Inc.                           \\
                                & Microsoft                     & IBM                               \\
                                & Dell                          & Intel                             \\
                                & Huawei                        &                                   \\
    \bottomrule
    \end{tabular}
}
\label{tbl:editoral_goldenset}
\end{table}

\subsection{Embedding Quality}

\subsubsection{Training Parameters}
We choose the following parameters for the lg2vec training process. For the input graph, we choose the combined graph (Top10-Combined-Wiki in Table~\ref{tbl:graph_dataset}) that comes with the top 10 Wikipedia languages, Wiki links, link-maintext graph, and clickstream graph as layers. Notice that in the clickstream graph layer, we flatten its edge weight distribution to be more linear by taking the weight to be $w^{0.75}$. For each layer of the graph, we sample these layered graphs with even importance by generating 50 random walks on each node with hop length from 5 to 20 in each layer. For word2vec training, we apply the skip-gram model with hidden dimension 200, window size 2-10, and 5 negative examples for each skip-gram pair. These embeddings are trained for up to 20 epochs. As demonstrated in Table~\ref{tbl:embedding_quality_parameters}, we observe that more epochs, longer hops for random walks or bigger window sizes only come with minor help for embedding quality. Therefore, in the remaining section, we adopt the setting of running 3 epochs, hop length 10 and windows size 3 for all experiments. For doc2vec, we set the window size to be 5.

\begin{table}[]
    \small
    \caption{Embedding quality with different training parameters for lg2vec. Accuracy of triplet evaluation on datasets in \cite{dai2014document} is reported.}
    \begin{tabular}{ccccc}
    \toprule
    Epoch & Hop & Window        & hand-built & generated \\ \midrule
    20    & 20   & 3            & 0.9705      & 0.8223   \\
    5     & 20   & 3            & 0.9588      & 0.8185   \\
    3     & 20   & 3            & 0.9647      & 0.8192   \\
    2     & 20   & 3            & 0.9647      & 0.8299   \\
    1     & 20   & 3            & 0.9647      & 0.8165   \\ \midrule
    3     & 15   & 3            & 0.9470      & 0.8223   \\
    3     & 10   & 3            & 0.9647      & 0.8124   \\
    3     & 5    & 3            & 0.9588      & 0.8192   \\ \midrule
    3     & 10   & 10           & 0.9411      & 0.8155   \\
    3     & 10   & 5            & 0.9529      & 0.8233   \\
    3     & 10   & 2            & 0.9294      & 0.8117   \\ \bottomrule
    \end{tabular}
    \label{tbl:embedding_quality_parameters}
\end{table}

\subsubsection{Layered Graph}

Table~\ref{tbl:embedding_quality_layers} presents the quality of the embeddings trained with different layers of the master graph. While the quality of the embedding trained with one single layer (En-Wiki-link) is already on par with the result of doc2vec, the embedding trained with multiple weight or language layers yields even better performance.

\begin{table}[]
    \small
    \caption{Embedding quality of different embeddings. Accuracy of triplet evaluation on datasets in \cite{dai2014document} is reported.}
    \begin{tabular}{lcc}
    \toprule
    Embeddings         & hand-built      & generated       \\ \midrule
    En-UniqLink-All    & 0.9352          & 0.7781          \\
    En-Link-Main       & 0.9647          & 0.7966          \\
    En-Clickstream     & 0.9647          & 0.8058          \\
    En-Master          & \textbf{0.9705} & 0.8223          \\
    Top10-Combined     & 0.9647          & \textbf{0.8329} \\ \midrule
    En-doc2vec         & 0.8869          & 0.7764          \\ \bottomrule
    \end{tabular}
    \label{tbl:embedding_quality_layers}
\end{table}

\subsection{Candidate Generation}

Here we demonstrate the candidate generation quality for each embedding by coverage. For a given query $q$ in a embedding, let the top-$k$ nearest neighbors of $q$ as $N_k(q)$ and the recommendations of $q$ from a golden set as $D_r$ 
the coverage for $q$ is defined as $|N_k(q) \cap D_r|/|D_r|$. Ideally, we would like an embedding to have high coverage with low $k$, which means the candidate set has better suggestion result even with fewer candidates provided. The candidate set's coverage is measured with three recommendation golden sets: Google's \emph{Person} and \emph{Company} recommendations and editorial's \emph{Company} recommendations.

\begin{figure}[htbp]
    \centering
    \includegraphics[width=0.9\columnwidth]{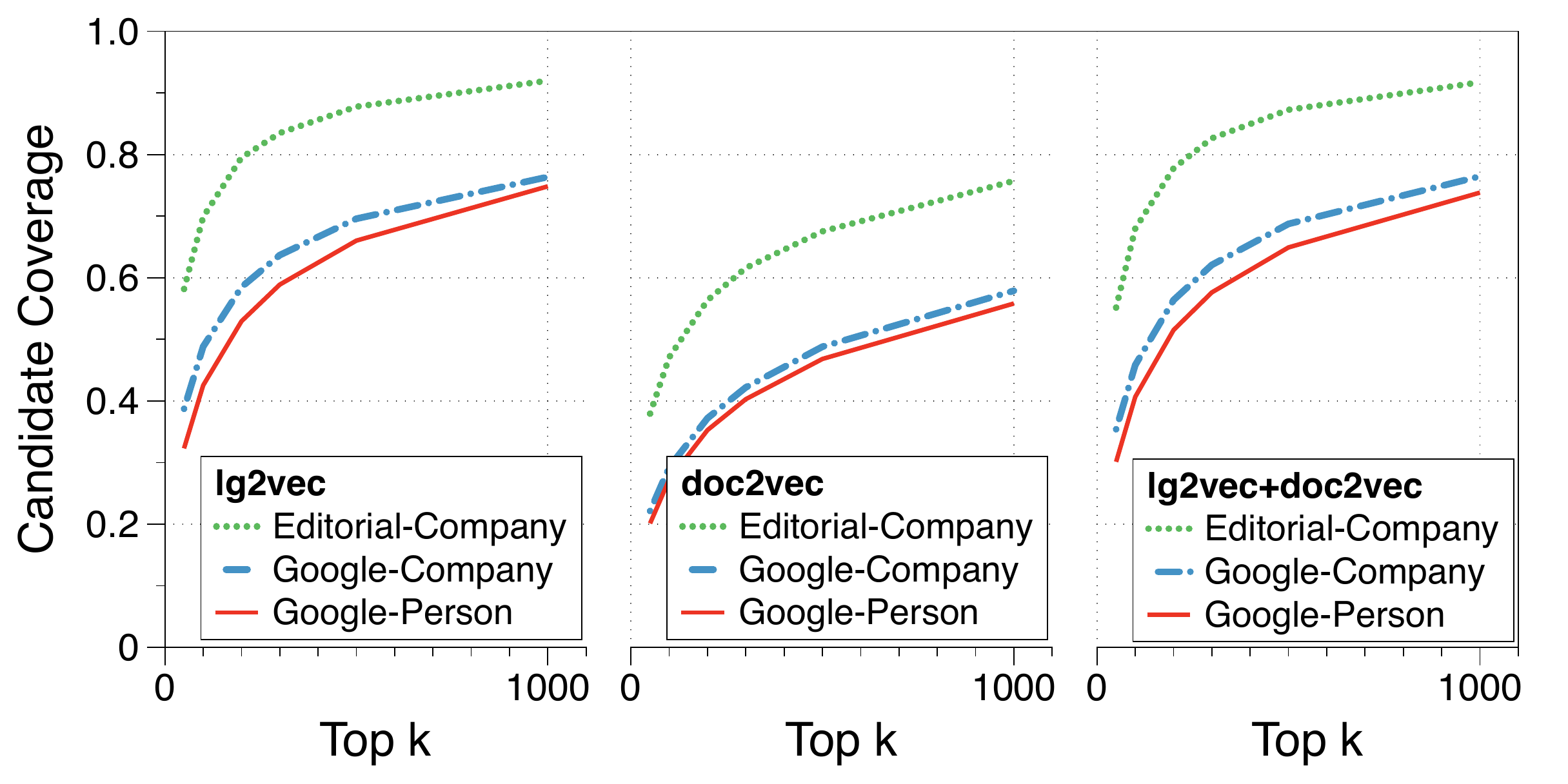}
    \caption{Coverage of candidates generated by top k nearest neighbors in different embeddings. 
    }
    \label{fig:candidate_coverage}
\end{figure}

Figure~\ref{fig:candidate_coverage} shows the coverage of candidate set suggested by top-$k$ nearest neighbors in lg2vec, doc2vec, and mixed suggestions from both lg2vec and doc2vec embeddings (lg2vec+doc2vec). In lg2vec+doc2vec, the top $k$ candidates are generated by mixing the top-$\frac{k}{2}$ nearest neighbors in lg2vec and doc2vec. The evaluation results denote that with $k=500$, lg2vec is up to $20\%$ better in coverage than doc2vec and almost on par with lg2vec+doc2vec. The reason behind lg2vec+doc2vec candidates is that we want to make sure to include candidates that capture both topological and semantic meaning, especially for those Wikipedia entities that do not include any links in their pages. In entity ranking, we rank the candidates generated from lg2vec+doc2vec with $k=500$.   



\subsection{Entity Ranking}

\subsubsection{Ranking Features}

\begin{table*}[]
    \caption{Evaluation result of related entity (Company) recommendation by rankers trained on Google Company / Person datasets. The average precision is computed using the relevant companies list curated by editors for 1012 queries of high pageviews companies. Our full model, which takes embeddings and other features, performs better than Google's related companies for MAP@6,9.}
    \centering
    \small
    \begin{tabular}{lcccccc}
        \toprule
        & \multicolumn{3}{c}{Company} & \multicolumn{3}{c}{Person} \\
        \cmidrule(r){2-4} \cmidrule(r){5-7}
        Features                                                & MAP@3  & MAP@6  & MAP@9  & MAP@3  & MAP@6  & MAP@9  \\
        \midrule
        lg2vec + clickstream                                    &     0.5168 &    {0.6509} &    {0.6723} & 0.5472 &    {0.6558} &    {0.6762}\\
        doc2vec + clickstream                                   &     0.4829 &    {0.5778} &    {0.5897} & 0.4939 &    {0.5860} &    {0.5906}\\
        lg2vec + doc2vec + clickstream (baseline)               &     0.5357 &    {0.6612} &    {0.6773} & 0.5493 &    {0.6544} &    {0.6742}\\
        baseline + pageview                                     &     0.5679 &    {0.6706} &    {0.6869} & 0.5959 &    {0.6843} &    {0.6986}\\
        baseline + pageview + popratio                          &     0.6016 &    {0.6932} &    {0.7026} & 0.6142 & \bf{0.6943} &    {0.7001}\\
        baseline + popratio                                     &     0.6035 &    {0.6871} &    {0.7010} & 0.6188 &    {0.6925} & \bf{0.7017}\\
        baseline + pageview + popratio + ysearchlog             &     0.6183 &    {0.6982} &    {0.7043} & 0.6133 &    {0.6846} &    {0.6988}\\
        \bf{baseline + popratio + ysearchlog}                   &     0.6259 & \bf{0.6989} & \bf{0.7052} & 0.6066 &    {0.6862} &    {0.6975}\\
        \midrule
        Google                                                  & \bf{0.6265}&      0.6570 &      0.6592 & \bf{0.6265} & 0.6570 & 0.6592\\
        \bottomrule
    \end{tabular}
    \label{tbl:eval_recommendation}
\end{table*}

For the supervised entity ranking, we apply the crawled \emph{Person} and \emph{Company} golden sets as the training set. We train the ranker with the pairwise loss approach in Learning to Rank model by XGBoost. For each entity in the golden set, we take the first $6$ recommendations as positive training pairs and randomly select $6$ entities from the end $100$ candidates of the candidate set (top-$400$ to top-$500$ candidates) as negative training pairs. 

The features we used for training pairs $(q,d)$ include the lg2vec and doc2vec cosine similarity of $(q,d)$, clickstream of link $d$ in page $q$, pageview of $d$, popularity ratio of $(q,d)$, and the search log co-occurrence from Yahoo! search of $(q,d)$. The popularity ratio (popratio) here is defined as $log\frac{pageview(d)}{pageview(q)}$ with the assumption that we only want to suggest candidate $d$ that is equal or more popular than the query entity $q$. The search log co-occurrences (ysearchlog) are Yahoo!'s in-house search signal that provides accumulated co-occurrence for \emph{Person} and \emph{Company} for search terms and sessions.

Table~\ref{tbl:eval_recommendation} shows the performance of rankers we trained with various combinations of features. We train our rankers with Google's \emph{Person} and \emph{Company} golden set and evaluate using Mean Average Precision at $k$ (MAP@$k$) with the editorial golden set. As a result, most of the rankers with lg2vec similarity as features perform better than Google in MAP@[6,9]. While the ranking results without the help of the ysearchlog already perform well, the ysearchlog features boost the MAP@3 for ranker trained with Google's \emph{Company}.

\begin{figure}[htbp]
    \centering
    \includegraphics[width=1\columnwidth]{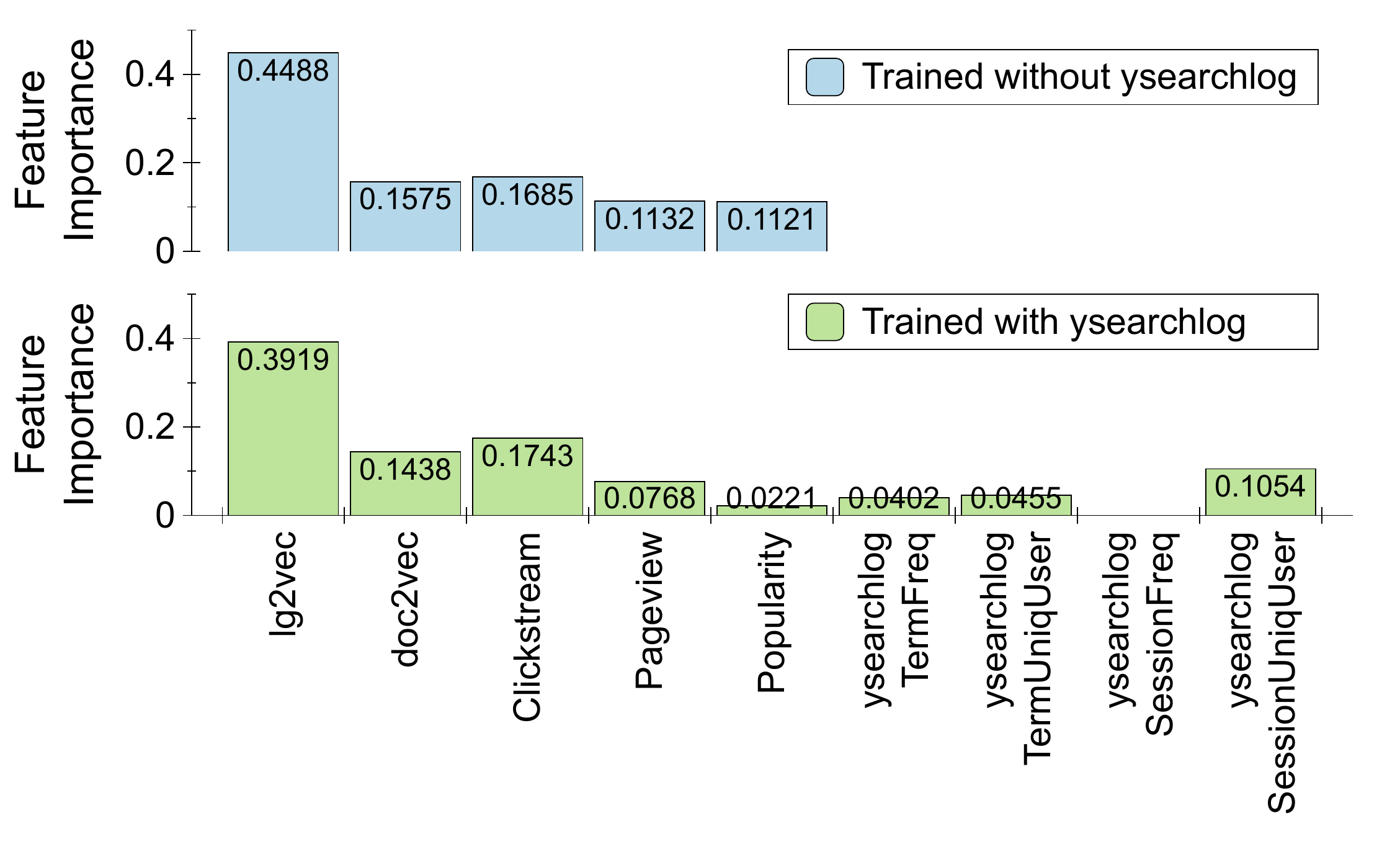}

    \caption{Feature importance of ranker trained with/without search log from Yahoo! Search US.}

    \label{fig:feature_importance}
\end{figure}

The feature importance of rankers computed by XGBoost using ``weight'' configuration is shown in Figure~\ref{fig:feature_importance}, in all cases, lg2vec similarity gives the highest feature importance for around $0.4$. It worth mentions that for Google \emph{Person} training set, we actually train our ranker with entities that are typed \emph{Person}, and evaluated with \emph{Company} typed editorial golden set. This shows that the features we selected are strong and general enough that it can support type agnostic recommendations. In the end, we choose the ranker with ``baseline + popratio + ysearchlog'' as our production ranker for related \emph{Company} recommendation in Yahoo! Search US.

\subsubsection{Entity Recommendation Results}

In Table~\ref{tbl:embedding_examples}, the entity recommendation results that are sorted by lg2vec's cosine similarity, doc2vec's cosine similarity and the ranker's score (baseline + popratio + ysearchlog) are presented. Since lg2vec is trained with the Wiki link graph's topology, lots of non-Person and List entities are included. In comparison, doc2vec recommends entities that have similar page contents in a semantic way. For the ranker, the entity recommendation results are already very promising, since we do not include any entity type features yet. 

For the final result, we filter the result with YK's entity type classifier to present the desired entity type. With the help of type filter, now it is possible to provide recommendations for the queries that do not have specific types (typed as \emph{Entity}) like ``Nonsteroidal Anti-Inflammatory Drug'', or for queries with specific type requirement such as to ask for the most related \emph{Person} to ``Apple Inc''. See Table~\ref{tbl:recommendation_examples} for more example queries with different query type filters.


\subsection{Web Search Evaluation}

To evaluate the performance of our entity recommendation model with real users, we created bucket tests on Yahoo! Search. Those bucket tests focused on measuring the performance of our related companies when users search for \emph{Company} entities. They were both held in August and September, 2019. Each bucket test was at least one week long and ran on $5\%$ of the traffic of Desktop version of Yahoo! Search US. Eventually, this experiment showed that our model brought a $1-3\%$ increase in click-through rate and resulted in about $5$ times more clicks per query session. Other tests will be conducted on other entity types.
\begin{table*}[]
\caption{Result of entity recommendation}
\small
\begin{tabular}{|c|C{2.7cm}|C{2.2cm}|C{2.2cm}|c|C{2.2cm}|c|}
\hline

Query entity     & \multicolumn{3}{c|}{Brad Pitt}                                                                                    & \multicolumn{3}{c|}{Yahoo!}                                \\ \hline

Sorted by        & lg2vec                                                             & doc2vec            & Ranker                  & lg2vec              & doc2vec              & Ranker       \\\hline \hline
1                & Brad Pitt filmography                                            & Angelina Jolie    & Angelina Jolie         & Yahoo! Search      & Timeline of Yahoo! & Google       \\\hline
2                & List of awards and nominations received by Brad Pitt       & Leonardo DiCaprio & Leonardo DiCaprio      & Verizon Media      & History of Yahoo!  & AOL          \\\hline
3                & Douglas Pitt                                                      & Jennifer Aniston  & Jennifer Aniston       & History of Yahoo! & Twitter              & Baidu        \\\hline
4                & Angelina Jolie                                                    & Ben Stiller       & Brad Pitt filmography & AOL                 & Google Search       & Yahoo! Mail \\\hline
5                & Satellite Award for Best Supporting Actor - Motion Picture & Patricia Arquette & Matt Damon             & Google              & Yahoo! Mail         & Facebook    \\\hline
6 & Matt Damon & Ben Affleck & Tom Cruise & Yahoo! Mail & Criticism of Google & Verizon Media \\ \hline
7 & Satellite Award for Best Actor - Motion Picture & Jennifer Garner & George Clooney & Baidu & LinkedIn & Yahoo! Answers \\ \hline
8 & Leonardo DiCaprio & Nicole Kidman & Julia Roberts & Timeline of Yahoo! & Criticism of Yahoo! & LinkedIn \\ \hline
9 & Jennifer Aniston & Reese Witherspoon & Sandra Bullock & Yahoo! Answers & Google & EBay \\ \hline
10 & Screen Actors Guild Award for Outstanding Performance by a Cast in a Motion Picture & Bradley Cooper & Robert Downey Jr. & Altaba & Tumblr & Web search engine \\ \hline
\end{tabular}
\label{tbl:embedding_examples}

\end{table*}

\begin{table*}[]
\caption{Type agnostic query entity recommendations}
\small
\begin{tabular}{|c|c|c|c|C{1.8cm}|C{2.2cm}|c|}
\hline

Query entity     & \multicolumn{3}{c|}{Apple Inc.}                 & Nonsteroidal Anti-Inflammatory Drug & World Marathon Majors & Me Too Movement   \\ \hline
Query type       & \multicolumn{3}{c|}{Company}                     & Entity                             & Sport Competitions  & Event             \\ \hline
Result type & Person         & Company            & Product    & Entity                             & Sport Competitions  & Person            \\ \hline \hline
1                & Steve Jobs    & Microsoft          & IPhone     & Paracetamol                        & Berlin Marathon    & Tarana Burke     \\ \hline
2                & Tim Cook      & Intel              & IPad       & Ibuprofen                          & Chicago Marathon   & Harvey Weinstein \\ \hline
3                & Steve Wozniak & NeXT               & MacOS      & Aspirin                            & London Marathon    & Rose McGowan     \\ \hline
4                & Jony Ive      & Google             & IOS        & Diclofenac                         & Tokyo Marathon     & Alyssa Milano    \\ \hline
5                & Ronald Wayne  & Beats Electronics & IPhone SE & Naproxen                           & Boston Marathon    & Uma Thurman      \\ \hline
\end{tabular}
\label{tbl:recommendation_examples}
\end{table*}
\section{Conclusion}
In this paper, we presented a layered-graph based embedding framework for entity recommendations based on public available Wikipedia dumps. The proposed lg2vec embedding can act as a standalone recommendation system but also as a supporting feature for any existing ranking pipeline. Our entity recommendation result is shown to perform well on editorial golden set and bucket tests on Yahoo! Search US. 

\bibliographystyle{ACM-Reference-Format}
\bibliography{kg}

\end{document}